\documentclass[3p,times,twocolumn,procedia]{elsarticle}
\usepackage{ecrc}

\usepackage[fleqn]{amsmath}
\volume{00}

\firstpage{1}

\journalname{Nuclear Physics B Proceedings Supplement}

\runauth{}

\jid{nuphbp}

\jnltitlelogo{Nuclear Physics B Proceedings Supplement}

\usepackage{amssymb}

\usepackage[figuresright]{rotating}

\begin{document}

\begin{frontmatter}




\title{\large Resonance Chiral Lagrangians and alternative approaches to hadronic tau decays}


\author{Pablo Roig}

\address{Instituto de F\'{\i}sica, Universidad Aut\'onoma de M\'exico, Apartado Postal 20-364, M\'exico D.F. 01000, M\'exico.\\
Departamento de F\'isica, Centro de Investigaci\'on y de Estudios Avanzados, Apartado Postal 14-740, 07000 M\'exico D.F., M\'exico.}

\begin{abstract}
Exclusive semi-leptonic decays of the tau lepton offer a clean probe to study the hadronization of QCD currents in its non-perturbative regime and 
learn about resonance dynamics, which drives strong interactions in these processes. In this theory outlook, I will use the simplest non-trivial di-pion 
tau decays to illustrate briefly recent theoretical progress on these analyses and their comparison to data.
\end{abstract}


\end{frontmatter}


\section{Introduction}
We will focus here on exclusive hadronic decays of the tau lepton. An updated detailed account on this topic, containing inclusive analyses, leptonic tau decays, 
and CPV and LFV searches in tau decays as well, can be found in Ref.~\cite{Pich:2013lsa}.\\
\indent The matrix element for $\tau^-\to H^- \nu_\tau$ decays, where $H$ stands for the final-state hadrons, can be written
\begin{equation}
 \mathcal{M}\left(\tau^-\to H^- \nu_\tau\right)\,=\,\frac{G_F}{\sqrt{2}}\,V_{ud/us}\, \bar{u}_{\nu_\tau}\, \gamma^\mu \,(1-\gamma_5)\, u_\tau\, \mathcal{H_\mu}\,,
\end{equation}
in which
\begin{equation}\label{H_mu}
 \mathcal{H_\mu}\,=\,\left\langle H \Big|(V-A)_\mu \,e^{i\mathcal{L}_{QCD}}\Big|0\right\rangle\,=\,\sum_i (...)^i_\mu\, F_i(q^2,...)\,
\end{equation}
is the hadronic matrix element of the left-handed QCD current evaluated between the initial hadronic vacuum and the final-state mesons. In eq.~(\ref{H_mu}), 
$(...)^i_\mu$ are the set of allowed Lorentz structures and $F_i(q^2,...)$ the hadronic form factors, scalar functions depending on the kinematical 
invariants ($q^2, ...$), which reduce to the charged-meson decay constants ($F_\pi, F_K$) for one-meson tau decays. This, in turn, are well-known from the 
measured $(\pi/K)^-\to\mu^-\bar{\nu}_\mu$ decay rates \cite{Agashe:2014kda}. Multi-meson modes start to provide non-trivial information on the hadronization 
of QCD currents.\\
\indent A model/theory is needed to compute the $F_i(q^2,...)$ in this case, and observables are readily obtained either directly in terms of them, or using their 
appropriate combinations, the structure functions \cite{Kuhn:1992nz}.\\
\indent $M_\tau\sim1.8$ GeV implies that the dynamics of hadronic tau decays will be mostly influenced by the lowest-lying light-flavored resonances like $\rho(770)$ or 
$a_1(1260)$, so that their propagation must be accounted for in the hadronic form factors.\\
\indent Among the many approaches that have been developed with this purpose, let us mention the Gounaris-Sakurai (GS) \cite{Gounaris:1968mw} and K\"uhn-Santamar\'{\i}a 
(KS) \cite{Kuhn:1990ad} parametrizations and the Resonance Chiral Lagrangians ($R\chi L$) approaches \cite{Ecker:1988te} that we will be discussing.\\
\indent These input form factors are fitted to data, directly or by means of a Monte Carlo Generator (which also helps to better estimate the backgrounds for a given 
process), being TAUOLA \cite{TAUOLA} the standard one in these low-energy applications. Related developments \cite{MCs} are of interest both for theorists and 
experimentalists, for flavor-factories and colliders.\\
\indent This kind of analyses should render the determination of resonance parameters: masses, widths and couplings. Contrary to the most common practice, model 
parameters should be avoided for the first two. Instead, physically meaningful model independent parameters shall be used, as they are those defined by the pole 
position of the resonance in the complex plane
\begin{equation}\label{pole}
 \sqrt{s^{pole}_{Res}}\,=\,M^{pole}_{Res}-\frac{i}{2}\Gamma^{pole}_{Res}\,,
\end{equation}
which should be the ones quoted by the PDG \cite{Agashe:2014kda}.\\
\indent ALEPH, CLEO, DELPHI and OPAL first \cite{pre-Bfacts}, and then the two-flavor factories -BaBar and Belle- \cite{Bfacts, Fujikawa:2008ma} have been providing data 
on exclusive hadronic decay modes of the tau lepton with increasing precision and the prospects for Belle-II \cite{Boris} and future planned facilities \cite{Future} 
are very much promising. This demands a corresponding effort on the theoretical description of these decays. Recent theoretical developments on hadronic tau 
decays are discussed in the next section using the well-known two-pion vector form factor (VFF) to explain them.
\section{An illustrative example: The two-pion VFF}
\begin{figure}[!t]
\begin{center}
\vspace*{0.2cm}
\includegraphics[scale=0.35]{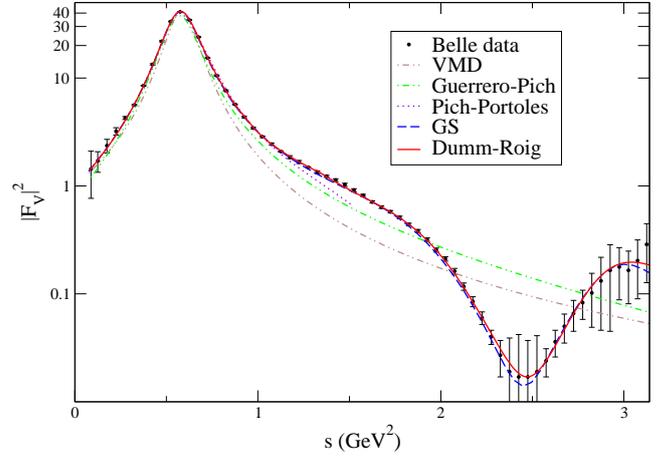}
\caption[]{\label{fig}
\small{Belle data \cite{Fujikawa:2008ma} (black dots) for $F_V(s)=F_V^{\pi^-\pi^0}(s)$ is compared to different theoretical descriptions: 
best agreement is obtained for our dispersive representation \cite{Dumm:2013zh} (solid), followed by the GS parametrization (dashed). The 
dispersive formula of Pich-Portol\'es (dotted) \cite{pipiRChT} agrees nicely with the data up to the $\rho'$ region, whose effect was not 
included in this parametrization. A similar feature is observed for the Guerrero-Pich (dotted-dashed) description \cite{pipiRChT}, which 
also restricted to $\rho(770)$ exchange. A na\"ive vector meson dominance result is also shown (dotted-dashed) to illustrate its departure 
from data at low energies as a consequence of violating the chiral limit at next-to-leading order.}}
\end{center}
\end{figure}

The KS-like parametrizations of the hadronic form factors are built fixing their normalizations to the leading order Chiral Perturbation Theory result \cite{Weinberg:1978kz}, 
as linear weighted combinations of Breit-Wigner factors accounting for the dominant exchanged resonances. For instance, in the case of the $\pi^-\pi^0$ vector 
form factor
\begin{equation}
 \left\langle \pi^-(p)\,\pi^0(p') \Big|\bar{d}\,\gamma_\mu\, u\Big|0\right\rangle\,=\,\sqrt{2}\,(p-p')_\mu\, F_V^{\pi^-\pi^0}(s),
\end{equation}
where $s=(p+p')^2$, the KS-parametrization reads
\begin{equation}
 F_V^{\pi^-\pi^0}(s)\,=\,\frac{BW_\rho(s)\,+\,\alpha\, BW_{\rho'}(s)\,+\,\beta \,BW_{\rho''}(s)}{1\,+\,\alpha\,+\,\beta}\,,
\end{equation}
with
\begin{equation}\label{BW_KS}
 BW^{KS}_R(s)\,=\,\frac{M_R^2}{M_R^2\,-\,s\,-\,i\,\sqrt{s}\,\Gamma_R(s)}\,,
\end{equation}
where the off-shell resonance width, $\Gamma_R(s)$, is obtained from the absorptive part of loop functions involving pions and Kaons. The GS expressions add to 
eq.(\ref{BW_KS}) a contribution resembling the one produced by the real part of these loops, shifting both numerator and denominator.\\
\indent However, both KS and GS parametrizations do violate the low-energy expansion of QCD \cite{ChPT} at next-to-leading order \cite{GomezDumm:2003ku}, introducing a 
systematic error in the analyses using them \footnote{Analyticity and UV QCD constraints are also violated \cite{Roig:2014mva}.}. This bias can (and 
should) be avoided by approaches built on the basis of chiral symmetry, as they are the $R\chi L$. In fact, these intend to interpolate between the two known 
extreme regimes of QCD: the chiral limit at low energies and the operator product expansion (OPE) of QCD at high energies ($E>M_\tau$). An extensive program 
\cite{Ecker:1988te, OPE-RChT} has been developed to work out the restrictions imposed on the resonance couplings by the OPE. The existence of a consistent 
minimal set of short-distance constraints applying to the two- and three-point Green functions and related form factors in the resonance region has been shown in 
both intrinsic parity sectors \cite{ConsistentSetSDConstraints}. These were obtained in the $N_C\to\infty$ limit \cite{largeN_C} within the single resonance 
approximation including multi-linear operators in resonance fields and working the latter in the antisymmetric tensor formalism. Such procedure provides a sound 
theoretical basis for the $R\chi L$ and their application to study two- and three-meson tau decays.\\
\indent A complementary approach uses dispersion relations to obtain the hadronic form factors. In this way, analyticity and unitarity are automatically fulfilled
and the poorest known (high-E) region is suppressed by the subtractions of the dispersive integrals. Consequently, results are also less sensitive to the precise
short-distance QCD constraints, minimizing the effect of the error associated to the $1/N_C$ expansion.\\
\indent Again, in the case of the charged two-pion VFF one would have \cite{Dumm:2013zh}
\begin{equation}\label{FV_3_subtractions}
 F_V^{\pi^-\pi^0}(s) \,=\,\exp \Biggl[ \alpha_1\, s\,+\,\frac{\alpha_2}{2}\,
s^2\,+\,\frac{s^3}{\pi}\! \int^\infty_{s_{\rm thr}}\!\!ds'\,
\frac{\delta_1^1(s')} {(s')^3(s'-s-i\epsilon)}\Biggr] \ ,
\end{equation}
for the three-subtractions \footnote{These are $\alpha_1$ and $\alpha_2$ -which are taken as free parameters- and $F_V^{\pi^-\pi^0}(0)=1$, which is fixed by CVC.} 
case, where the phaseshift, $\delta_1^1(s)$ is obtained as
\begin{equation}\label{delta}
 \tan \delta_1^1(s) = \frac{\Im m \left[F_V^{\pi^-\pi^0(0)}(s)\right]}{\Re e \left[F_V^{\pi^-\pi^0(0)}(s)\right]} \ ,
\end{equation}
using an input form factor $F_V^{\pi^-\pi^0(0)}(s)$, which can be provided by the $R\chi L$ \cite{pipiRChT}
\begin{eqnarray} \label{SU2formula}
& & \hspace{-.5cm} F_V^{\pi^-\pi^0\,(0)}(s) = \frac{M_\rho^2}{M_\rho^2 \left[1+\frac{s}{96\pi^2 F_\pi^2}\left(A_\pi(s)+
\frac12 A_K(s)\right)\right]-s} \\
& & = \frac{M_\rho^2}{M_\rho^2 \left[1+\frac{s}{96\pi^2 F_\pi^2}\Re e \left(A_\pi(s) + \frac12 A_K(s)\right)\right]-s-i M_\rho \Gamma_\rho(s)}\ .\nonumber
\end{eqnarray}
\indent In eq.~(\ref{SU2formula}) the whole complex chiral loop functions $A_P(s)$ \cite{ChPT} have been resummed in the denominator ensuring analyticity. Unitarity 
is also warranted in the elastic region. Beyond it, inelastic coupled channel effects need to be included. These are essential in obtaining the scalar 
$(K\pi)^-$ \cite{Jamin:2001zq} and $\pi^-\eta$ form factors \cite{DeltaS=0}. For final states with more than two mesons, only Ref.~\cite{Moussallam:2007qc} has 
taken into account these effects in the $(K\pi\pi)^-$ form factors.\\
\indent Constructs analogous to eqs.~(\ref{FV_3_subtractions}-\ref{SU2formula}) have been employed also in the $(K\pi)^-$ VFF \cite{KpiVFF} and first attempts 
at describing the $K^-\eta$ VFF have also been undertaken \cite{Keta}. In particular, a joint analysis of the $\tau^{-} \to K_{S} \pi^{-} \nu_{\tau}$ 
and $\tau^{-} \to K^{-} \eta\nu_{\tau}$ decays has recently allowed \cite{Joint} an improved determination of the $K^*(1410)$ pole mass and width parameters 
according to eq.(\ref{pole}) \footnote{For resonances overlapping with neighboring states, such as the lightest axial-vector resonance with strangeness 
$K_1(1270)$, this pole determination might not be feasible.}
\begin{equation}
 \hspace{-.22cm} M_{K^{*'}}^{pole}\,=\,(1304\pm17)\,\mathrm{MeV},\;\Gamma_{K^{*'}}^{pole}\,=(171\pm62)\,\mathrm{MeV}\,.
\end{equation}
\indent The analysis of Belle data \cite{Fujikawa:2008ma} for the number of events distribution measured in 
$\tau^-\to\pi^-\pi^0\nu_\tau$ decays yields the pole position \cite{Dumm:2013zh}
\begin{equation}
 M_\rho^{pole}\,=\,(761\pm2)\,\mathrm{MeV},\;\Gamma_\rho^{pole}\,=(142\pm2)\,\mathrm{MeV}\,,
\end{equation}
in nice agreement with alternative determinations.\\
Two-pion VFF Belle data \cite{Fujikawa:2008ma} are compared to different theoretical approaches in fig.~\ref{fig}.
\section{Multi-meson modes and radiative corrections}
Three-meson tau decays provide a much richer dynamical structure with up to four participating form factors. A reasonably good understanding has only been 
achieved for the three-pion modes \cite{GomezDumm:2003ku, 3pi} but ongoing efforts extend also to the $(KK\pi)^-$ \cite{Dumm:2009kj}, $\eta\pi^-\pi^0$ \cite{Dumm:2012vb} 
and $(K\pi\pi)^-$ \cite{Kpipi} decay channels. A description of the state-of-the-art parametrizations for the $(\pi\pi\pi)^-$ and $(KK\pi)^-$ hadronic form factors 
in TAUOLA is given in Olga's talk \cite{MCs}.\\
\indent For multi-meson modes the biggest challenge is to go beyond the $R\chi L$ description of $\tau^-\to(\pi\pi\pi\pi)^-\nu_\tau$ of Ref.~\cite{Ecker:2002cw}, 
including operators multi-linear in resonance fields, specially taking into account the soon expected release of Belle data on these decay spectra \cite{Hayashii}. 
For even higher-multiplicity modes, there are only parametrizations based on the KS model \cite{Kuhn:2006nw} and isospin relations \cite{iso}, which provide a 
first estimate on these decays. More exotic modes have also been studied recently \cite{Othermodes}.\\
\indent For the most abundant decay modes radiative corrections become an issue. Consequently, they have been studied for the one-meson \cite{1mrad} and $\pi^-\pi^0$ 
\cite{2pirad} and $(K\pi)^-$ modes \cite{Kpirad} (see Ref.~\cite{GLC}) where they are important for testing lepton universality, extracting $V_{us}$ and obtaining 
$a_\mu^{HVP,LO}$, respectively, using tau decay data. These corrections are still lacking for the $(\pi\pi\pi)^-$ decay modes where they will be certainly needed 
with the advent of Belle-II data.
\section{Conclusions and outlook}
Hadronic decays of the $\tau$ lepton are a clean probe of the hadronization of QCD currents in the light-flavor sector at low energies. While inclusive studies 
are ideal for extracting fundamental parameters like $\alpha_S$ or $V_{us}$, exclusive analyses are able to determine resonances properties with precision.\\
\indent $R\chi L$ and dispersion relations are the best theoretical approaches to deal with hadronization in these decays. Two-meson and three-pion decay modes are well 
understood, so the challenge is now on controlling the corresponding radiative corrections.\\
\indent All other multi-meson modes are still not described satisfactorily. Dedicated effort is needed from the collaboration between experiment and theory through 
Monte Carlo Generators, specially with forthcoming flavor facilities in mind.
\section*{Acknowledgements}
I thank and congratulate Ian M. Nugent and Achim Stahl for the excellent organization of this Workshop. 
This work is partly funded by the Mexican Government through CONACYT and DGAPA (PAPIIT IN106913).







\end{document}